# High Performance Computing Evaluation
# A methodology based on Scientific Application Requirements

Mariza Ferro, Antonio R. Mury, Laion F. Manfroi, Bruno Schlze

National Laboratory of Scientific Computing,

Getulio Vargas 333, Petropolis, Rio de Janeiro

**Abstract**:
High Performance Distributed Computing is essential to boost scientific progress in many areas of science and to efficiently deploy a number of complex scientific applications. These applications have different characteristics that require distinct computational resources too. In this work we propose a systematic performance evaluation methodology. The focus of our methodology begins on scientific application characteristics, and then considers how these characteristics interact with the problem size, with the programming language and finally with a specific computational architecture. The computational experiments developed highlight this model of evaluation and indicate that optimal performance is found when we evaluate a combination of application class, program language, problem size and architecture model.

## 1. Introduction

Scientific computing involves the construction of mathematical models and numerical solution techniques to solve complex scientific and engineering problems. These models often require a huge processing capacity in computer resources to perform large scale experiments within a reasonable time frame. These needs have been addressed with High Performance Parallel and Distributed Computing (HPDC), which allows many scientific domains leverage progress. However, it is very difficult for many research groups to evaluate these HPDC infrastructures and arrive at the best configuration to run their scientific applications.

Usually, optimal configurations are searched by executing one of the existing benchmark suites, widely used for performance evaluation. Benchmarks are good for comparisons between computational architectures, but they are not the best approach for evaluating if an architecture is adequate for a set of scientific applications. Evaluations using traditional benchmarks, return a single performance number that corresponds to, for example, the maximum number of floating-point operations per second. In contrast with those applications which typically are floating-point intensive, many scientific applications do not corresponds to this model, and even the workload often used. In other words, traditional benchmark evaluations generally, don't consider the actual set of applications that will be used. However, each application has different system requirements (e.g., memory bound, I/O bound and CPU bound) and so it requires different computational resources.

Thus, within the performance evaluation methodology proposed in this work, in order to achieve adequate performance evaluation it is necessary first to consider the characteristic of the scientific application that will be used in the HPDC architecture, under conditions as real as possible. In this way, the parameters evaluated differ from those usually evaluated when the focus is performance optimization. The parameters evaluated, that we refer to here as Essential Elements of Analysis (EEA), are application's class, execution time, programming language, problem size/workload, average memory time and percentage of memory, CPU and I/O usage; in contrast with Flops/s, cache miss rate and cache hit rate.

The methodology under development comprises several phases and dozens of steps that enable researchers to evaluate which is the best HPDC configuration for their scientific applications set. The development of our methodology is rooted in two concepts: the first one is Operational Analysis (OA) [1], which is the foundational basis for our methodology. OA involves a sequence of phases and steps that aim to determine the performance of a system under the most realistic operational conditions. The second one is the Dwarfs of scientific computation, developed by Colella [2] and Berkeley team [3], that enable the application requirement characterization. Each Dwarf class characterizes applications by common requirements in terms of computation and data movement.

Although the methodology under development comprises several phases and steps, in this work we briefly describe the overall methodology, describing in detail one of these steps with the experimental setup that enabled its development. The experimental results highlight how different interactions among the EEA, such as application's class and computational architectures, can deliver performance results that are completely diverse. The proposed methodology is supported by these results, and it is presented in the next sections of this work.

The remainder of the paper is organized as follows: In Section 2 we discuss the scientific landscape and how applications can be categorized in classes using Dwarf taxonomy. In Section 3 we discuss related work. In Section 4 we discuss the traditional performance evaluation paradigm followed by our proposal for performance evaluation. Section 5 outlines our experimental setup and results. Section 6 concludes de paper and briefly discusses future work.

## 2. Applications and Dwarfs

With the aim of categorizing the styles of computation seen in scientific computing, the work of Colella [2] identified seven numerical methods that he believed to be important for science and engineering and introduced the "Seven Dwarfs" of scientific computing. These Dwarfs are defined at a high level of abstraction to explain their behavior across different HPDC applications, and each class of Dwarfs shows similarities in computation and communication. According to his definition, applications of a particular class can be implemented differently with the change in numerical methods over time, but the underlying patterns have remained the same over generations of change and will remain the same in future implementations. These dwarfs were neither particular software applications nor were they small benchmark kernels. Instead, they represented entire families of computation with common computational properties.

The Berkeley team in parallel computation extended this classification to thirteen Dwarfs after they examined important application domains. They were interested in applying Dwarfs to a broader number of computational methods and investigating how well the Dwarfs could capture computation and communication patterns for a large range of applications. Ideally, the Berkeley team would like good performance across the set of Dwarfs to indicate that new manycore architectures and programming models will perform well for a broad range of future applications. Traditionally, applications target existing hardware and programming models but instead, they wanted to design hardware keeping future applications in mind [4]. They compared the Dwarf classes against collections of benchmarks for embedded computing (42 benchmarks EEMBC - http://www.eembc.org/) and for desktop and server computing (28 benchmarks SPEC2006 - http://www.spec.org/cpu2006). Additionally, they examined important application domains: artificial intelligence/machine learning, database software and computer graphics/games. The goal was to delineate application requirements to draw broader conclusions about hardware requirements. A diverse set of important scientific applications is supported by the current 13 Dwarfs. A more complete discussion about this can be found in [4, 5, 6, 3].

The Dwarf classes under investigation in this present work are Dense Linear Algebra (DLA) and Graph Traversal (GT). We focus on these classes because there are many scientific applications in many scientific areas classified in these classes. Additionally, we focus on these classes because these two together offer a diverse set of patterns that comprises a more complete set of experiments. Scientific applications classified in the DLA class are computationally bound while in the GT class are memory bound. However, to investigate whether one single application completely captures the breadth of a Dwarf, our longer term investigations will include more than one application for DLA class, which could present different aspects of a given Dwarf. Next, Dwarf classes are described along with the applications used in this work.

**1. Dense Linear Algebra** Dwarf class computations involve this set of mathematical operators performed on scalars, vectors or matrices when most of the matrix or vector elements are non-zeros. Dense in this Dwarf refers to the data structure accessed during the computation. The arithmetic intensity of the computation operating upon the data may be low intensity operators (scalar-vector, vector-vector, matrix-vector, matrix-matrix, vector reduction, vector scan and dot product) that carry a constant number of arithmetic operations per data element. This Dwarf has a high ratio of math-to-load operations and a high degree of data interdependency between threads. These set of mathematical operators are the basis of more sophisticated solvers such as LU Decomposition (LUD) or Cholesky and exhibit high arithmetic intensity [5]. Generally, such applications use unit-stride memory accesses to read data from rows and strided access to read data from columns. Applications classified as DLA are relevant across a variety of domains such as in material science to molecular physics and nanoscale science; in energy assurance for combustion, fusion and nuclear energy; in fundamental science such as astrophysics and nuclear physics; in engineering design for aerodynamics. Representative algorithms of this class are LUD, matrix transpose, triangular solver, symmetric eigensolver, clustering algorithms such as Kmeans and Stream Cluster, and many others. We performed experiments with the LUD and Kmeans algorithms.

(a) **LUD** is an algorithm to calculate the solutions of a set of linear equations that decomposes a matrix as the product of a lower triangular matrix and an upper triangular matrix to achieve a triangular form that can be used to solve a system of linear equations easily. A ma- trix $A \in R^{n \times n}$ has a LU factorization iff all of its leading principal minors are non-zeros, i.e., $det(A[1:k,1:k]) \neq 0$ for $k = 1 : n-1$.

(b) The **Kmeans** is a well-known clustering algorithm used extensively in data mining. It is a method that partitions n points that lie in $d-$dimensional space into k clusters in this way: seeded with k initial cluster centers, Kmeans assigns every data point to its closest center, and then recomputes the new centers as the means of their assigned points. This process of assigning data points and readjusting centers is repeated until it stabilizes.

**2. The Graph Traversal** Dwarf class applications must traverse a number of objects in a graph and examine characteristics of those objects such as would be used for search. A graph or a network is an intuitive and useful abstraction for analyzing relational data where unique entities are represented as vertices, and the interactions between them are depicted as edges. The vertices and edges can further be assigned attributes based on the information they encapsulate. Such algorithms typically involve a significant amount of random memory access for indirect lookups and little computation [5]. Scientific domains that include important applications in this class and examples of application are bioinformatics (MuMMer), graphs and search (Breadth-First Search and B+Tree).

(a) **B+Tree** is an *n*-ary tree often with a large number of children per node. A B+Tree consists of a root, internal nodes and leaves. The root may be either a leaf or a node with two or more children. The primary value of a B+Tree is in storing data for efficient retrieval in a block-oriented storage context because B+Trees have very high fan-out (number of pointers to child nodes in a node, typically on the order of 100 or more), which reduces the number of I/O operations required to find an element in the tree. The order, or branching factor, b of a B+Tree measures the capacity of nodes (i.e., the number of children nodes) for internal nodes in the tree. The actual number of children for a node, referred to here as m, is constrained for internal nodes so that $[b/2] \leq m \leq b$. Leaf nodes have no children, but are constrained so that the number of keys must be at least $[b/2]$ and at most $b-1$. In the situation where a B+Tree is nearly empty, it contains only one node, which is a leaf node. The root is also the single leaf in this case. This node is permitted to have as little as one key if necessary and at most b.

Some evidence for the existence of the equivalence classes proposed by the Dwarfs can also be found in some numerical libraries such as The Fastest Fourier Transform in the West (FFTW) [7], a software library for computing discrete Fourier transforms (equivalent of Spectral Methods Dwarf class), the LAPACK/ScaLapack [8] software library for numerical linear algebra (equivalent of DLA Dwarf class) and OSKI [9], a collection of primitives that provide automatically tuned computational kernels on sparse matrices (equivalent of Sparse Linear Algebra class - SLA). The thirteen Dwarfs are also related to the Intel classification of computation in three categories: Recognition, Mining and Synthesis (RMS). The RMS applications are considered important to guide new architectural research and development that comprises applications in Artificial Intelligence and Machine Learning, databases, games and computer graphics. These applications are represented by diverse Dwarf classes, such as DLA, SLA, Spectral Methods, Backtrack and Branch Bound, and others [4]. Rodinia [10], Parboil [11], Torch [5] and Parallel Dwarfs Project (http://paralleldwarfs.codeplex.com/) are open-source benchmark suites that implement applications based on a subset of the 13 Dwarfs.

These examples motivate the Dwarf use as a way of categorizing scientific applications, both for the importance of libraries and application areas mentioned as well as for these recent benchmark suite developments, which cover new architectures and could indicate the relevance and contemporariness of these classes for the scientific community.

The experiments conducted in this work using Dwarf classes are presented in section 5.

## 3. Related Work

In this section, we briefly review the related work that have any connection with our proposal. We first review research based on the interaction that exists between applications characteristics and performance. We then review works using Dwarfs as base of development and afterwards some works using OA, in the context of computational performance evaluation.

3.1. Scientific Applications Characterization

There are many efforts to seek a more reliable way to evaluate the performance of computing systems. Among these efforts, many have focused on the importance of these to be directed to your set of scientific applications. Some examples of application domain-specific benchmark suites have been proposed, like MediaBench[12], CommBench [13] and BioBench [14].

There are some projects that are going in the same direction to search for a more reliable measure of system performance to their scientific applications. Similar to our proposal, the projects Mantevo [15] and CORAL [16] are focus- ing the performance evaluation based on the set of applications. The Mantevo project is focused on developing tools to accelerate and improve the design of high performance computers and applications by providing application and library proxies to the high performance computing community. These applications, called Miniapps, represent a class of application and the performance- intensive aspects of the application with the idea that, although an application may have one million or more lines of source code, performance is often dom- inated by a very small subset of lines. Extracting a few hundred lines of code is much easier to port and to test several software performance questions and ideas.

That project is similar to ours in the sense they believe that application performance is determined by a combination of many choices: hardware platform, runtime environment, languages and compilers used, algorithm choice and implementation. In this complicated environment, they use Miniapps for exploring the parameter space of all these choices. Although Miniapps share similar ideas with Dwarfs (even some of them are similar with Dwarf classes), there are only four Miniapps. Further, Mantevo focus is to understand an application to allow its optimization, while our focus are the researchers in any domain of application. They often do not want to change their code, but to get the best computer to run their applications instead. However, knowing the critical computational requirement of the application, nothing prevents a team to work on that aspect of optimizing code.

The CORAL project is a collaboration of Oak Ridge National Laboratory, Argonne National Laboratory and Lawrence Livermore National Laboratory referred to as the Laboratories, for three pre-exascale High Performance Computing (HPC) systems to be delivered in the 2017 timeframe. They intend to choose two different system architectures and procure a total of three systems. The project describes specific technical requirements related to both the hardware and software capabilities of the desired system as well as application requirements. The application requirements are represented by a set of representatives benchmarks aimed at exploiting performance features of the CORAL systems. The CORAL benchmarks have thus been carefully chosen and developed to represent the broad range of applications expected to dominate the science and mission deliverables on the CORAL systems. Moreover, there is an incisive interest in protecting their investment in the DOE application base by procuring systems that allow today's workhorse application codes to continue to run without radical refacturing.

Their proposal is very similar to ours, since they intend to evaluate performance based on their set of applications and for this purpose they provide a set of benchmarks representing their application set's requirements. Besides, this set of benchmarks must meet minimum performance measurements, similar to Measures of Effectiveness (MOE) in our methodology. The differences lie in the main objective. While our focus is to evaluate architectures currently available in an attempt to satisfy applications requirements, they want to develop a new one. Additionally, the representative set of applications is proprietary and selected by ad-hoc knowledge, while ours is representative of a more generalized knowledge.

*3.2. Dwarfs Characterization*

In particular, the Dwarfs characterization of applications is important for this work and with this focus are the following work related. Since Phillip Collela in his 2004 presentation [2] gave his list of Seven Dwarfs to categorize the styles of computation seen in scientific computing, some researches were developed applying this concept. Berkeley work extended the concept for thirteen Dwarfs [4]. These two works is more detailed in Section 2.

The work [17] analyses workloads with more complex data movement pat- terns and discusses changes on architectural requirements in the context of these workloads. They discuss a data-centric workload taxonomy that seeks to separate the most important dimensions across which these applications differ. By examining existing and emerging workloads, they argue for a systematic approach to derive a coverage set of workloads based on this taxonomy, inspired by Dwarfs.

The work of [6] focuses on GPU implementations of some selected Dwarfs and discusses three benchmark suites which implement a subset of the 13 Dwarfs on the GPU. They list typical problems related to efficient GPU implementations and discuss the specific problems and performance with respect to some GPU Dwarfs.

The Torch project [5] identified several kernels for benchmarking purposes in the context of high-performance computing. They argue that a number of existing benchmarks can be seen as reference implementations of one or more kernels from TORCH. The kernels are classified according to the 13 Dwarfs and authors discuss possible code optimization strategies that can be applied to these. For each Dwarf, several algorithms are included in the suite which are different in the implementation detail, but all of them are part of a higher level Dwarf.

The Parallel Dwarfs project [18] teams adopt the Berkeley's 13 Dwarfs classification to describe the underlying computation in each of their benchmarks. It corresponds to a suite of 13 kernels parallelized using various technologies such as OpenMP, TPL and MPI code.

Rodinia [10] and Parboil [11] are open source benchmark suites which implement applications that were mapped to a subset of the 13 Dwarfs. The Rodinia applications are designed for heterogeneous computing infrastructures, and uses OpenMP and CUDA to allow comparisons between manycore GPUs vs. multi-core CPUs. The Parboil's implementations are on GPU and some basic CPU implementations.

Some works propose scientific application characterization to improve performance on cloud computing environment. Examples using Dwarfs to predict performance are [19], Hawk-i [20] and [21]. Cloud computing was not the focus of this work, but our methodology is capable of being applied to cloud computing and was initially evaluated for cloud in [22] and evaluated how different Dwarf classes interact in virtualized environment.

All the aforementioned works are very close to ours in the sense of highlighting the importance of characterizing scientific applications to better understand the available infrastructure and many of them use Dwarfs for that. Furthermore, some works mapped a set of benchmarks to Dwarfs classes, such as [4], [10] and [11], using them to evaluate performance in high performance architecture. However, none of them developed any model to know which Dwarf class an application corresponds to, neither describe any methodology of how to apply their concepts in the evaluation of architectures, nor evaluated the behavior of Dwarfs under different architectures.

*3.3. Operational Analysis*

OA has been used with great successes in the military decisions since the advent of World War II and also spread extensively to nonmilitary applications after that. So, OA emerges as successful method also in government and in industry. In computation was firstly proposed by [23] and has been used for computational performance analysis for prediction as to assist in understanding the performance of applications. Other work was developed for Queuing Networks [24], [25] and [26].

The paper [27] applies operational analysis to the problem of understanding and predicting application-level performance in parallel servers. They present operational laws that offer both insight and actionable information based on lightweight passive external observations of black-box applications. The Occupancy Law accurately estimates queueing delays and enables improved monitoring and system management. The Capacity Adjustment Laws bound the performance consequences of capacity changes in a real parallel network server, enabling capacity planning and dynamic resource provisioning.

The work [28] presents a pair of performance laws that bound the change in aggregate job queueing time that results when the processor speed changes in a parallel computing system. The results show that OA usefully complements existing parallel performance analysis techniques.

In the work [29], the computational model developed, using OA and other theories, employs an alternative characterization of uncertainty that is expressed entirely in terms of observable phenomena. This alternative characterization provides significant benefits: it leads to solutions that are directly applicable to practical problems, it provides a new perspective on the analysis of risk, and it supports a new method for improving the efficiency of certain computer driven simulations.

# 4. Performance Evaluation Methodology

The design of high performance supercomputers is on the boundary of constant and significant changes. With the arrival of petascale computing in 2008, we see another potential paradigm shift in the construction of parallel computing and the use of hybrid designs employing heterogeneous computational accelerators. In these designs, specialized hardware devices such as graphics processing units (GPUs), dense multi-core processor, the forthcoming Knights- range of many-integrated core (MIC) and field programmable gate arrays (FPGAs) have gained significant interest throughout the HPC community. Both of which represent significant concerns in the design of petascale and exascale supercomputers in the coming decade. However, despite the impressive theoretical peak performance of accelerator designs, several hardware/software development challenges must be met before high levels of sustained performance can be achieved by HPC codes on these architectures.

According to [30] it is very common for users of HPC systems to be dis- appointed with the actual performance of their hardware. Among the main reasons for this is that the conventional way to evaluate an architecture is using a benchmark suite based on existing programs. But, the benchmark essentially delivers a performance number that uniquely characterizes it relative to other systems, without considering attributes from an application's perspective. If the benchmark model does not capture the important aspects of target machines and applications, then the analysis is not predictive of real performance. For ex- ample, the Linpack benchmark [31] is the most widely recognized and discussed metric for ranking HPC systems. However, Linpack *"is increasingly unreliable as a true measure of system performance for a growing collection of important science and engineering applications"* [32]. The dominant calculation in this algorithm are dense matrix-matrix multiplication and this kind of algorithm strongly favors computers with very high floating-point computation rates and adequate streaming memory systems. In contrast to these, many applications operate with distinct operation behavior, such as, for example bioinformatics application as analysis presented in [14].

Benchmarks are an useful way to evaluate performance when we know what technology to experiment with and what application or workload it represents. To address this challenge, we need benchmarks that capture the execution pat- terns (i.e., dwarfs) of applications, both present and future, in order to guide performance evaluation. Further, while there is a lot of work on how to design good algorithms for these highly-threaded machines, in addition to a significant body of work on performance analysis, there are no systematic theoretical models to analyze the performance of programs on these machines. We need a systematic model to analyze the performance of applications on these machines. In this sense we propose a systematic methodology with which performance evaluation focused on scientific applications is possible, as well as finding the most adequate computational architecture for them. This methodology is presented next.

*4.1. Proposed Methodology*

In this section we briefly describe the proposed methodology for HPDC performance evaluation. The main focus is on the set of scientific applications and not on the architecture. The main objective of this methodology is to assist researchers and technical staff to determine what is the best architectural design/equipment for their application set. Because for those people, although HPDC is essential, they don't want become HPDC specialists and ultimately choose the architecture by maximum TFlops/s or core count, which may not be the best choices. This choice must be guided by the set of applications that will run on this architecture.

With this situation in mind, our methodology was developed, as mentioned, using two basics concepts as reference: Dwarfs and OA. The Dwarf classification was chosen due to the need to characterize scientific applications, since they are the focus of our evaluation. Towards this goal we use the Dwarf application pattern to write our ideas down and put them into a systematic form that can be used by others. Motivations and details were explained in Section 2.

From Operational Analysis we extracted a model to evaluate complex systems. Its objective is to be systematic in the analysis of possible actions to provide a decision-making process for rationally choosing an architecture. The OA method involves the use of scientific methods to bring objectivity to the results, and to make verification possible. The methods are quantitative, using techniques of mathematics and other sciences to deal with quantitative aspects of a problem [33]. The OA method attempts to determine the performance of a system under the most realistic operational conditions.

The OA is conducted to estimate the system's utility, operational effective- ness, and operational suitability, and need for any modification. OA is a dynamic process, since the operational environment, is continually changing and OA must therefore be extended over the entire life cycle of a system [1].

Motivated by OA historical and background in the military scenario, where contributions to evaluate their highly complex systems were exceptional, we developed our method facing HPDC systems, using as a reference OA method. The methodology under development is described next.

*4.1.1. Description of methodology*

In order to reach the aforementioned objectives, we are developing a systematic methodology conducted in a set of sequential phases and steps. The complete set of phases and their corresponding steps is presented in Figure 1; in each phase a report is developed for an appropriate decision point. However, in this work we will briefly explain the whole diagram, and choose to focus instead on step "MOEs & EEAs" and its respective experiments.

The first phase is the Definition Problem in which the real problem must be defined and OA objective clearly defined. Finding the real problem and real objectives is one of the most difficult hitches. For example, a common mistake is to determine the problem as obtaining a new HPDC system; however, the real problem is to execute a set of scientific applications with a reasonable performance. So, a clear definition of objectives and specific purposes related to the process of OA, will be critical to plan and to implement the other phases of the OA. The prerequisites (infrastructure, costs and time) are initially defined, enabling a initial modeling elaboration and a guideline under which the process is conducted. Also, its scope, constraints and resources are defined. Schedule and preliminary evaluation report are elaborate. The Problem Detailing Analysis phase detail the user problem searching complete requirements definition (implicit, explicit and tacit requirements).

Very important here is the knowledge acquired about each application in the scientific workflow: the real problem sizes/workload executed, programming languages, applications executed sequentially or in parallel, etc. Further, the applications are mapped to a Dwarf class[1] and an impact for each one in the workflow. Beyond that, critical issues, EEA (predefined or new ones defined exclusively for attempt user's requirement) and Measures of Effectiveness (MOEs) are defined. A MOE of a system is a parameter that evaluates the capability of the system to accomplish its assigned goal under a given set of conditions. They are important

---

[1] The model for mapping applications to Dwarf class was developed for this methodology, based on a set of experiments and using Machine Learning methods. But is a subject for further work.

because they determine how test results will be judged.

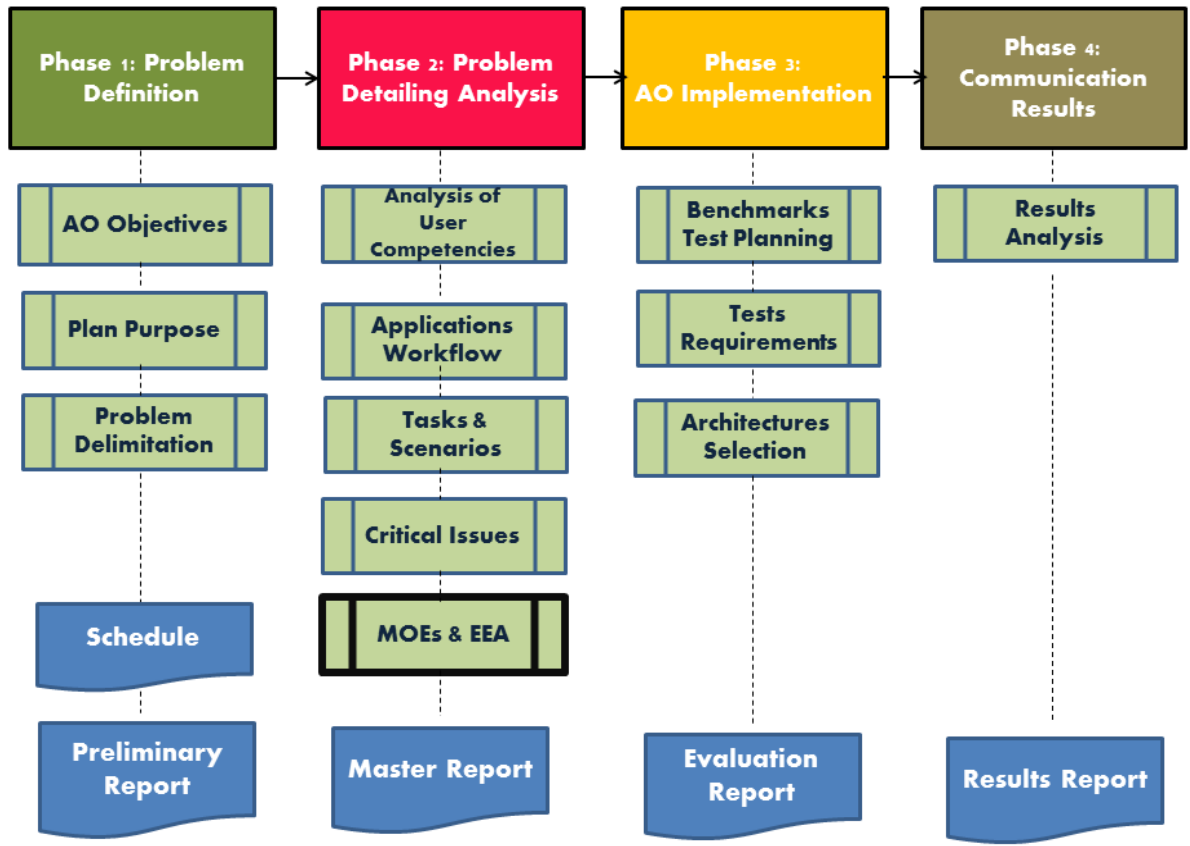

Figure 1: Proposed methodology for performance evaluation. Sequence of phases and their steps.

Within the OA implementation phase, the test planning is completed, based on both aforementioned phases. For each application class, one or more benchmarks are defined to be executed[2] and predefined EEA collect. Each EEA are categorized as mandatory, recommended or optional. A set of test procedures and a data collection model is generated to be provided to suppliers. In addition, the architectures are selected to be tested based on three criteria: suitability, feasibility and acceptability. The evaluation report is elaborated.

The last phase is communication of results, in which data collected are confronted with MOEs and the data from different providers are compared. Using MOEs and equations defined for this methodology it is possible to define the operational effectiveness and suitability. All final results are reported.

---

[2] The mapping from Dwarf class to benchmark was defined for this methodology, based on our experiments and theoretical material available. But by space constraints will be not presented here.

The methodology must therefore be extended over the entire life cycle of a system to evaluate changes, and to reevaluate the system to ensure that it continuously meets operational needs and retains its effectiveness, thus extending its life cycle and adapting it for of new requirements.

The methodology is yet under development, and should be improved as new classes of applications are studied, all those phases and steps are detailed in a User Manual. The Manual for OA applied to HPDC that we developed contains a detailed model for application, models to mappings applications to Dwarfs class and from Dwarf class to benchmark, and all details necessary for its application.

Next the EEA are presented and in Section 5 the experiments that led to this set of EEA.

*4.2. Essential Elements of Analysis*

The EEAs are a set of parameters evaluated in our methodology. Experiments (presented in Section 5) highlight the importance of these for a real performance evaluation, where the primary focus is on the set of scientific applications, with secondary emphasis on the methodology remaining as current as possible, so less focused on specific architecture design evaluation or performance code optimization. There are a set of predetermined EEAs in our methodology, but others can be defined at the time of methodology application. The predetermined EEAs are:

- Total Execution Time - the time between the start and the completion of an task; the latency to complete a task, including disk accesses, memory accesses, input/output activities and operating system overhead;
- Class of Application - Dwarf's class;
- Problem Size/workload;
- Programming language - the programming language coding;
- I/O Time - the time waiting for input/output (I/O) operations;
- Memory access time - the time that the processor waits for the memory;
- CPU access Time - the time the processor is computing, not including the time waiting for I/O or executing other programs.

Memory Access time, CPU time and I/O time are measured in absolute time (milliseconds) and percentage time (the percentage of CPU, I/O or memory time from the total execution time)[3].

The question could be, why don't measure another parameter, like cache L1, L2 (and who knows L3 ?) miss rate, GFlops/s, memory latency, among others more conventional. The answer is because, as we mentioned, our focus is not performance optimization or architecture evaluation (and for that those parameters are useful), but the application and users. Users usually don't want/can optimize their scientific codes, they just want to acquire (or upgrade) a HPDC system to execute their set of scientific applications with a reasonable performance. For that they will evaluate the better infrastructure as a black-box.

However, if users want to optimize their scientific code, our method can be useful too, given that its format captures the most essential elements, in such a way that a software designer can quickly find what he needs to solve a problem. When the code section bounding the application performance is known, it is possible to attack it. For example, if the methodology identifies the application as a GT class, we know that it is memory bound, or has percentage of memory access time higher then percentage of CPU time and I/O. Moreover, the Memory access time, is defined as:

$$Memory\ access\ time = Hit\ time + miss\ rate + miss\ penalty$$

In that case the user could measure Memory access time, not in a black box mode as we do, but they could measure cache hit time, miss rate and miss penalty, using traditional benchmarks for that. According to Hennessy [34], the Average Memory access time is a better measure than miss rate and this formula above give us three metrics for cache optimizations, by reducing hit time, miss rate and miss penalty.

## 5. Methodology and Experimental Setup

The main objective of the experiments is to capture the most essential elements of a problem's solution in such a way that we could choose EEA to our methodology and the best

---
[3] Network Bandwith access time and its percentage of use when using MPI implementation is another EEA and is currently under investigation, which is not ready for presenting results here.

solution (architectural needs) to solve a problem (optimize performance for a set of applications).

Different experiments were carried out to assess the influence of the applications' classes in computing results. The first group of experiments aimed to identify key influences of computer architecture versus the applications' classes. In addition to that we aimed to identify the effect of the type of programming language used in Dwarfs' implementation and also the workload size. In all cases different architectures were confronted.

*5.1. Selected Dwarfs*

The experiments presented in this work were conducted using two Dwarfs and varying datasets size as inputs. For the DLA Dwarf class LUD and Kmeans algorithms were used (Section 2). For the DLA experiments we also aimed to verify the consistency of Dwarf classes. The experiment results validate the categorization, and even though LUD and Kmeas are quite different algorithms they presented very similar computational requirements[4]. GT Dwarf class was tested using B+Tree algorithm. Details of the algorithms and motivations have been presented in Section 2.

For testing we use various problem size as input for the algorithms. For LUD experiments we use ten different matrix sizes, ranging from $2048 \times 2048$ up to $32768 \times 32768$. For Kmeans we use thirteen datasets, from size1 (1638400 objects) to size13 (9830400 objects). For B+Tree we use graph datasets from 2M nodes to 50M nodes.

Those three algorithms are available on Rodinia Benchmark suite [10] based on Berkeley's Dwarf. For the tests was used the default Rodinia's implementation, without any special setting up in the code for the processor and accelerator architectures. Ensuring the execution of the same code was significant, avoiding differences that may occur due to setting up the code better for one platform than for another.

*5.2. Selected Libraries*

All experiments also sought to determine the effect of the type of libraries used in the implementation of the Dwarfs. For testing OpenMP [35] and OpenCL [36] libraries were used because both allow parallel execution on a CPU, OpenCL was also used in GPU.

In OpenMP, programmers enable parallel execution by annotating sequential codes with "pragmas". Sequential algorithms are parallelized incrementally, and without major restructuring. The parallelism granularity in OpenMP can be controlled manually by adjusting the number of OpenMP threads in combi- nation with a scheduling type. An OpenCL program has two parts: compute kernels (executed on one or more OpenCL devices), and a host program that manages kernels execution.

---

[4] The details of experiments are not presented here, since it is not the focus of this work.

For OpenMP experiments, the number of threads for each test was specified to allocate all available cores. We also specify "scatter" as the thread affinity for this work, as specified in [37].

### 5.3. Selected Hardware Plataforms

The experimental infrastructure used three architectures, summarized in Table 1.

|  | **X86 based Multi-core (Arch A)** | **x86 based Multi-core (Arch B)** | **Manycore GPU Based (Arch C)** |
|---|---|---|---|
| **Theoretical Peak Performance** | 1177 GFlops | 281,6 GFlops | 1030 GFlops |
| **Memory Bandwidth** | --- | --- | 148 GB/s |
| **Memory Clock** | --- | --- | 148 GB/s |
| **Cores** | 64 | 32 | 448 |
| **Clock (GHz)** | 2.3 | 2.2 | 1.15 |

Table 1: Target Architectures used in this work.

We are not disclosing the commercial brands of CPU architectures used because the objective of this work is not to evaluate and compare performance from different manufacturers, but how their characteristics impact on the results instead.

### 5.4. Performance Comparison and Analysis

In each experiment presented next, 30 runs were made for each point and the average and standard deviation calculated. The confidence interval for the tests was less than 1%, so they are omitted in the graphs. A logarithmic scale was used for all graphs presented in this Section.

**Application's Class versus Architectures** Figure 2 and 3 show the results with two Dwarf's classes (DLA and GT) and its respective algorithms (LUD and B+Tree) both implemented in the same program language (OpenMP). These experiments were all conducted on the same architecture (Arch A) and for ten problem sizes. For each problem size, presented in ascending order, were made 30 runs. In the graphs are presented for each problem size the average percentage of resource utilization for each resource (CPU, I/O and memory).

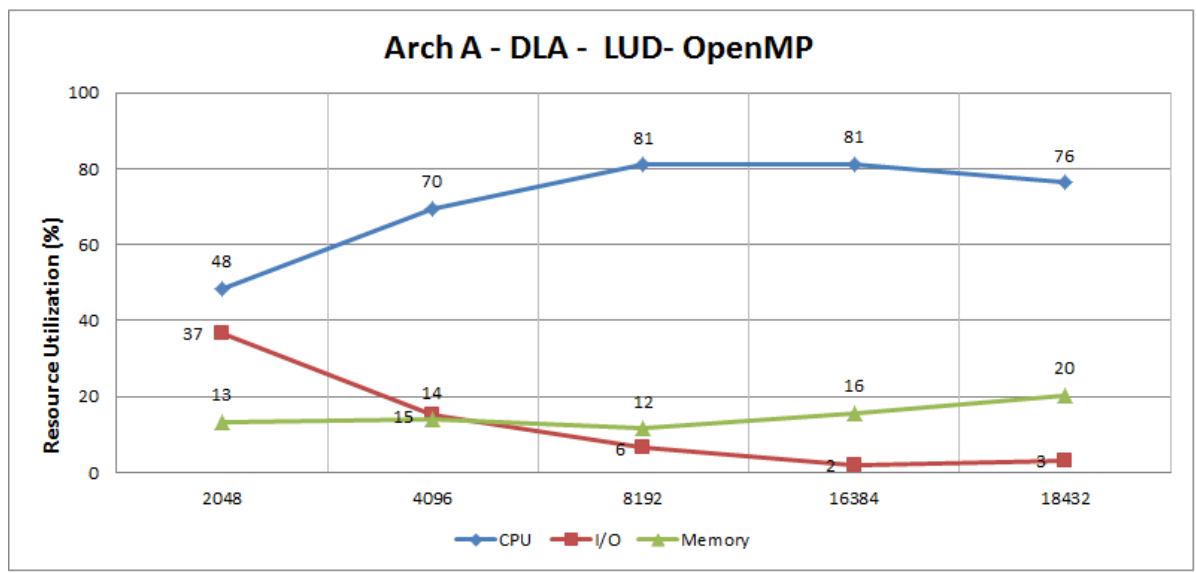

Figure 2: Graph of percentage of resource utilization for LUD application algorithm implemented in OpenMP and running in Arch A.

It's possible to note that for each Dwarf class, differences in terms of resource consumption are well characterized. While in the same architecture LUD was totally CPU bound (Figure 2) for all problem sizes, the main resource consumption for B+Tree is initially I/O and, as the problem size increases, B+Tree becomes memory bound. The relative consumption to CPU is negligible for B+Tree (Figure 3).

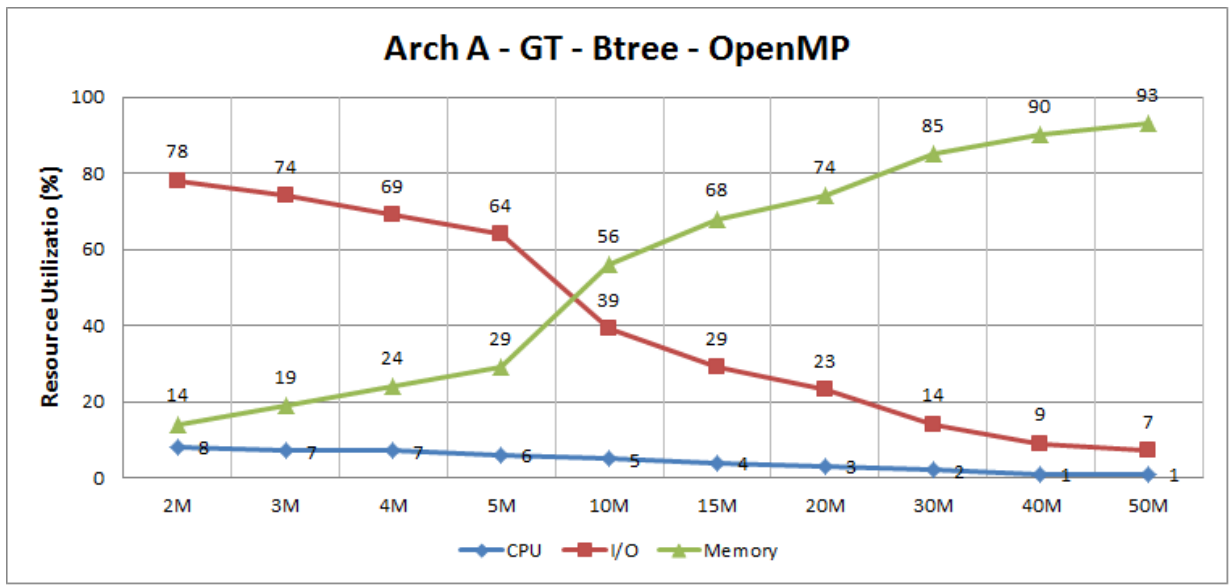

Figure 3: Graph of percentage of resource utilization for B+tree application algorithm implemented in OpenMP and running in Arch A.

It's clear the computational resource requirements differences to the same architecture, the same program language, but for different application class. These behaviors demonstrated the significance of knowing the scientific application and make an architecture acquisition oriented for that. The same behavior was observed when the Dwarf classes were executed in other architectures and other programming language. However, because of limited space graphics are not displayed in this work.

These experiments, besides validating the classes behavior, also verify and validate using the class of application as an EEA. This EEA proves important to characterize application resource requirement and thus to evaluate real performance. Knowing the application category, for example, the right hardware can be better selected. The next experiments highlight more that relationship.

**Application's Class versus Architectures and Program Language** Figures 4 and 5 show the performance as total execution time (in milliseconds - ms) when running DLA Dwarf class represented by Kmeans algorithm. The graphs show experiments conducted on thirteen problem sizes, presented in ascending order, and each of then run 30 times. The same algorithm was implemented in both OpenMP and OpenCL to verify a possible influence of programming language on performance.

In the Figure 4 is presented Kmeans performance implemented in OpenMP. Kmeans was executed in Arch A and Arch B and it is possible to note that for all problem sizes Arch A achieved better performance than Arch B. The total execution time in Arch A is approximately two times lower than in Arch B and in some sizes reaches three times. The experiments presented in Figure 5 were conducted in the same fashion as in the pre- vious scenario, also including Arch C because the OpenCL programming language make this possible. However, the performance for Kmeans im- plemented in OpenCL was very different. In this case in any execution Arch A had the best performance, but now Arch B is a better option for this algorithm

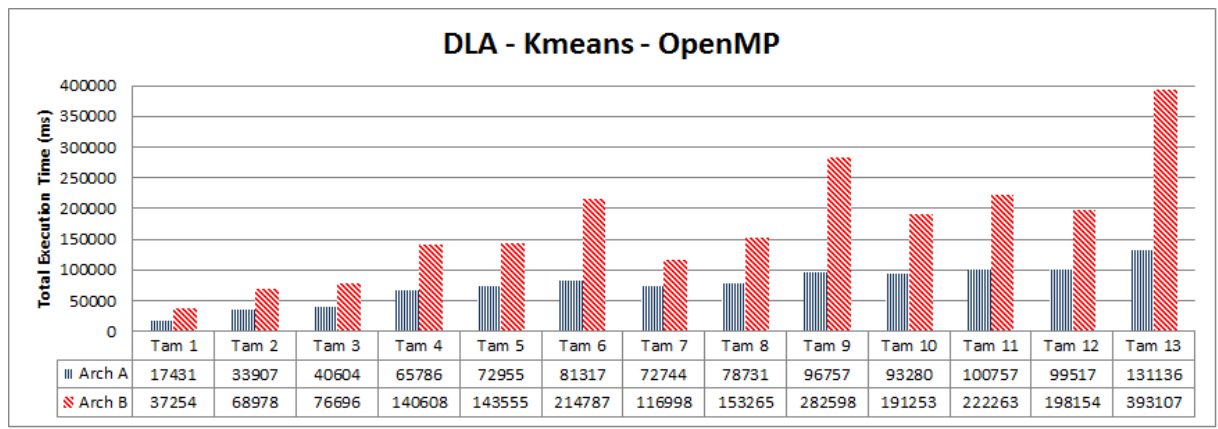

Figure 4: Graph performance of average total execution time for Kmeans application algorithm implemented in OpenMP and running in Arch A and Arch B.

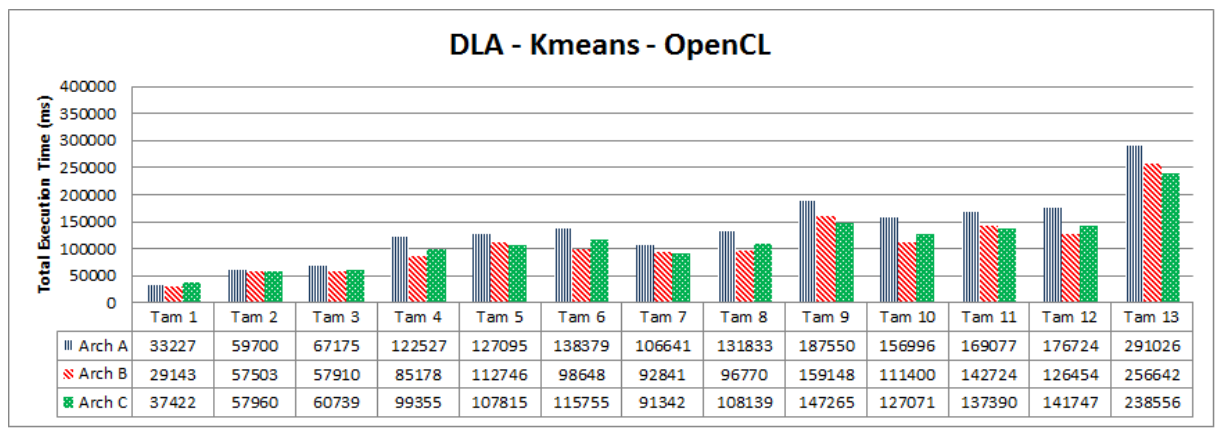

Figure 5: Graph performance of total execution for Kmeans application algorithm implemented in OpenCL and running in Arch A, Arch B and Arch C.

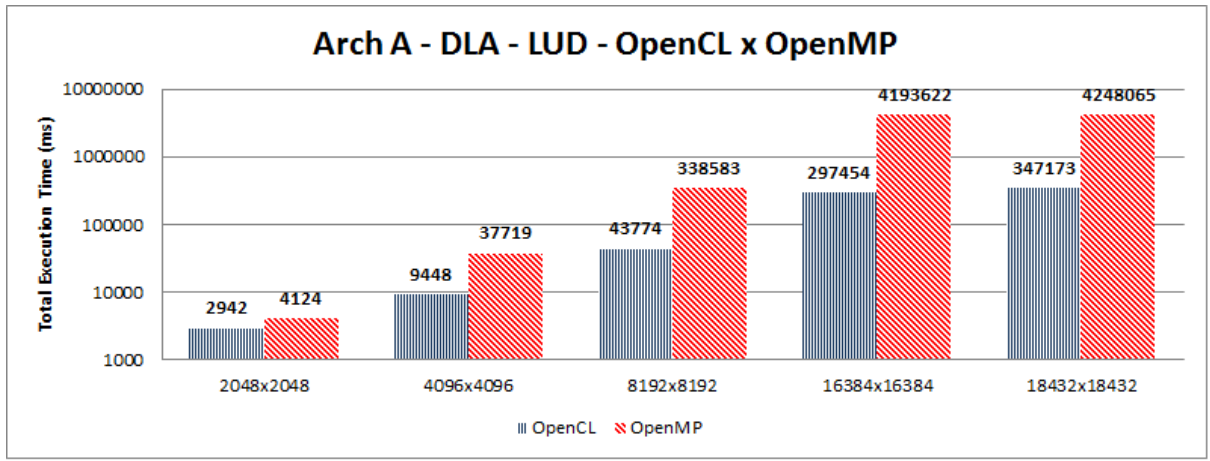

Figure 6: Graph performance of average execution time for LUD application algorithm implemented in OpenCL and OpenMP and running in Arch A.

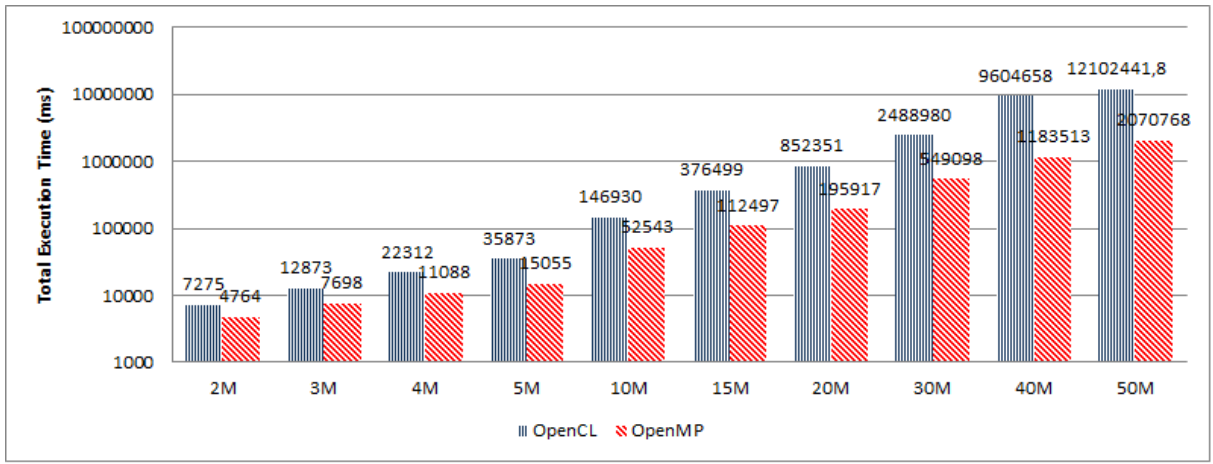

Figure 7: Graph performance of average execution time for B+tree application algorithm implemented in OpenCL and OpenMP and running in Arch A.

The influence of pairing programming language and architectures turned out clear after these experiments. Performance results change completely when the same applications' class was implemented in a different programming language. The program language influence was so significant that it changed the relative consumption of resources. While Kmeans implemented in OpenMP was CPU bound, the OpenCL implementation was memory bound. This relationship could even change the class of an application (these behaviors did not occur for the other algorithms tested).

In addition, in the Figure 6 and 7, we can observe other situations that confirm program language as relevant EEA for performance evaluation. In the Figure 6 LUD was obtained by running (30 runs) in Arch A, comparing both implementations (OpenCL and OpenMP). It can be seen that per- formance, expressed as average execution time, for OpenCL is consistently higher than for OpenMP implementation. The average execution time for OpenCL is at least two times lower, and as the problem size increases the time differences became even greater. The same behavior occurs in Arch B (graphs not presented here).

However, for the application class GT, represented by B+Tree algorithm, also in Arch A and comparing both implementations (Figure 7), OpenMP implementation performance is very different. In this case OpenMP has worse performance for all problem sizes. OpenCL is at least two times faster than OpenMP with significantly higher ratios as the problem size increases. The same behavior occurs in Arch B (graphs not presented here).

| **Alg/Class** | **Arch** | **OpenMP** | **OpenCL** |
|---|---|---|---|
| LUD/DLA | A | 4248065 | 347173 ↗ |
| LUD/DLA | B | 5138300 | 340379 ↗ |
| Kmeans/DLA | A | 70850 ↗ | 143206 |
| Kmeans/DLA | B | 219644 | 121150 ↗ |
| B+Tree/GT | A | 2070768 ↗ | 12102442 |
| B+Tree/GT | B | 639871 ↗ | 5102661 |

Table 2: Summary of experiments using OpenMP and OpenCL and the average execution time (ms) is presented for the largest input in the diverse set of architectures and algorithms.

These results are summarized in Table 2 where experiments using OpenCL and OpenMP are faced with diverse set of architectures (column Arch) and algorithms with respectively class (column Alg/Class). The implementation with better performance is indicated with ↗ in these combination and the average execution time (ms) is presented for the largest input.

These results point out the importance in evaluating the application class versus architecture and programming language, since OpenMP and OpenCL are so different in their approach to parallelism. Although this paper is focused neither on understanding the causes that lead to these divergent performance differences, nor on evaluating which is better (OpenCL or OpenMP), some reasons could be pointed out:

The possible cause that leads to this differences in performance can be explained in the class of application and the better cache utilization on OpenMP. So, since B+Tree is memory intensive, this could lead to a much better performance and OpenMP outperforms OpenCL in both architectures. But, we can see that OpenCL is a good choice for LUD, since memory is not the bottleneck, but CPU. However, this does not repeated for Kmeans, probably because what was already pointed out, that for this algorithm the programming language alters the application's behavior.

Comparative studies between OpenMP and OpenCL that focus on code optimization for specific platforms and for Dwarfs algorithms can be found in [38], [39] and [40].

The results presented in Table 2 highlight the importance regarding the actual programming language used in application when evaluating performance. It's important to remember that many times when acquiring a new architecture, the researchers do not intend to change their application code. In many situations, by doing this change, years of development and working hours would be lost. Furthermore, the results also point out that evaluating performance by executing a set of benchmark is unsatisfactory when this does not represent the real application requirements.

**Problem Size** Another EEA that became important while experiments were performed was the problem size. It's important to consider the actual problem size that is utilized for correctly evaluating performance. Some architectures versus class of applications may exhibit performance tracks. Example for that is presented in Figure 8 when LUD application was running in Arch A and Arch B. Experiments for this algorithm were performed using different problem sizes, with matrix sizes varying from 2048x2048 to 32768x32768. It's possible to note that when running LUD up to matrix size 16384x16384, the performance in Arch B is better than in Arch A. However, from matrix size 18432x18432 up to 28672x28672 the perfor- mance trend changes and Arch A becomes better. Even more so, from 32768x32768 the trend is reversed again.

These results show how architectures could present performance tracks according to problem size and application class used. The Arch B outper- forms Arch A up to input size 18K, when Arch B reaches its maximum performance capacity for 32 cores. Then, Arch A outperforms Arch B up to input size 32K, when Arch A also reaches its maximum performance for 64 cores. From this input size onwards, both have already achieved their maximum performance, and Arch B returns to be better, when comparing both architectures in their full capacity for this Dwarf class.

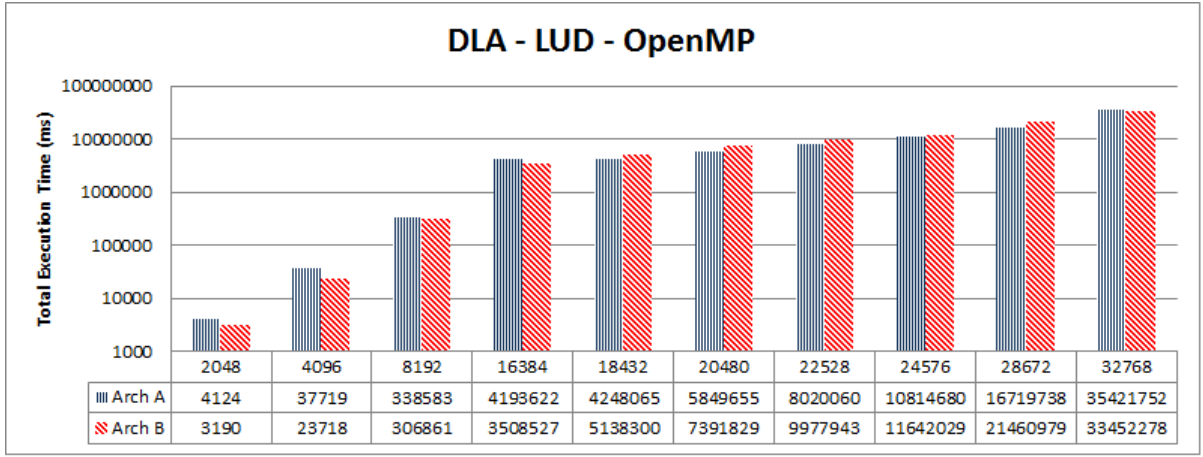

Figure 8: Performance in average execution time for LUD algorithm running in Arch A and Arch X for ten problem sizes (matrix from 2048x2048 to 32768x32768).

## 6. Final Considerations and Future Work

In this paper we presented a performance evaluation paradigm that changes the current and most common paradigm of using benchmarks suites. Our results showed that evaluating performance may not be so simple as to investigate a maximum peak performance obtained when executing a benchmark suite on an architecture. Our proposal was neither to evaluate which is the best architecture nor how to optimize one. Also, the proposal presented here was not about which program language is better for parallel computing. Instead, our proposal is primarily how to determine performance requirements for a set of scientific application.

But, that is a complex task and the importance of this study increases as the parallel computing revolution has been presented and a myriad of new computing architectures, promising higher performances, appears on average every six months. With the main objective being the evaluation of performance, at- tempting all these features (application class, programming language, workload size, etc.), we proposed a systematic methodology in which these features are investigated together. The methodology allows to identify which features are essential to evaluate performance of an application, here named as Essential Elements of Analysis.

Among the EEA that we proposed, first to understand which characteristics the applications have and to achieve this we used the Dwarf class characterization. Computation and communication patterns of these Dwarfs lead to diversified execution behaviors, presented in experimental results, thus corroborating the suitability of the Dwarf concept as a means to characterize which computer architectures perform better in face of different types of scientific applications. Understanding which type of applications perform well makes it easier to decide when to use one architecture or another. The right hardware can be better selected when you know the right job category. So don't just buy a more ex- pensive option when you can start with a cheaper one. Along the same line, determining the maximum performance is easier to compare within a group of algorithms.

Different experiments were carried out to assess the influence of applications' classes in computing results. The first group of experiments aimed to identify key influences of computer architecture versus the applications' classes. Added to that was aimed identify the effect of the type of program language used in Dwarfs' implementation and also the workload size. Always confronting with the different architectures. The results showed that both the programming language and workload size significantly influence performance, the latter with varying effect as a function of load ranges. These results highlight that the evaluation of performance of scientific applications comprises a combination of real application requirements (here represented by Dwarf classes), actual workload size, programming language and system architecture.

As future work, we will continue to evaluate other Dwarf classes. Addition- ally, exploring experiments considering the EEA for average time and percentage of use of bandwidth network. As such, results with MPI will also be analyzed.

**Acknowledgment**

The authors would like to thank the financial support from the Coordination for the Improvement of Higher Education Personnel (CAPES), the National Counsel for Technological and Scientific Research (CNPq) and the Funding Agency of the State of Rio de Janeiro (FAPERJ) for the support of this project.